\documentclass[aps,prl,reprint,superscriptaddress,showpacs]{revtex4-1}
\usepackage{graphicx}

\usepackage{amsmath}
\usepackage{amsfonts}
\usepackage{amssymb}

\usepackage{bm}
\usepackage{sidecap}

\usepackage{epstopdf}
\usepackage{hyperref}

\begin{document}

\title{Planar Metamaterial with Toroidal Moment}

\author{Yuancheng Fan}
\affiliation{Key Laboratory of Advanced Micro-structure Materials (MOE) and Department of Physics, Tongji University, Shanghai 200092, China}
\affiliation{Ames Laboratory and Department of Physics and Astronomy, Iowa State University, Ames, Iowa 50011, USA}
\author{Zeyong Wei}
\affiliation{Key Laboratory of Advanced Micro-structure Materials (MOE) and Department of Physics, Tongji University, Shanghai 200092, China}
\author{Hongqiang Li}
\email[Electronic mail: ]{hqlee@tongji.edu.cn}
\affiliation{Key Laboratory of Advanced Micro-structure Materials (MOE) and Department of Physics, Tongji University, Shanghai 200092, China}
\author{Hong Chen}
\affiliation{Key Laboratory of Advanced Micro-structure Materials (MOE) and Department of Physics, Tongji University, Shanghai 200092, China}
\author{Costas M. Soukoulis}
\affiliation{Ames Laboratory and Department of Physics and Astronomy, Iowa State University, Ames, Iowa 50011, USA}

\date{\today}

\begin{abstract}
We observe toroidal response from a planar metamaterial comprised of asymmetric split ring resonators (ASRRs). It is showed that a toroidal-molecule can be constructed through rational arrangement of planar ASRRs as meta-atoms via manipulating structural symmetry of the meta-atoms. Field maps clearly indicate that the toroidal resonance paves new electromagnetic confinement style in a subwavelength scale. Planar scheme of manipulating the coupling among the ASRRs may stimulate research in optical region involving toroidal multipoles. Toroidal geometry together with the Fano resonance made high-Q response will have enormous potential applications in low-threshold lasing, cavity quantum electrodynamics and nonlinear processing.
\end{abstract}


\maketitle
Toroidal moment is of fundamental interest in the physics community \cite{1}. The toroidal response produced by currents flowing on the surface of a torus along its meridians is firstly considered by Zel'dovich to explain parity violation in the weak interaction \cite{2}. The so called anapoles \cite{2} have gained acceptance in nuclear and particle physics \cite{3,4}. Theoretical calculation has also predicted toroidal dipole in a kind of molecular structure \cite{5}. Toroidal multipoles are different from electric or magnetic multipoles in traditional multipole expansions. Particularly, non-radiating feature of the toroidal geometry is quite unique which comes from its fantastical field localization configuration in a head-to-tail manner. However, the scattering of external fields from toroidal multipoles in nature is often much weaker for observation. So, suppressing traditional electric and magnetic multipoles and enhancing toroidal response by rationally manipulating structural symmetry are basic and essential towards a toroidal geometry.

Since there exist profound analog between the weak and the electromagnetic interactions, if we can reconstruct this unusual interaction in manipulating electromagnetic wave which is important in our modern life? Metamaterial is just a concept of artificially constructing micro-structured materials with fancied properties unattainable in nature, these artificial materials usually consist of subwavelength-sized metallic resonant building blocks as meta-atoms \cite{6,7}, which yield electric or/and magnetic responses. The local resonances of the meta-atoms and mutual coupling among them enable the manipulation of electromagnetic waves at a subwavelength scale.

Toroidal metamaterial was first theoretically proposed by Marinov et al. in 2007 in investigating a 3D-array of toroidal solenoids \cite{8}, later Papasimakis et al. experimentally studied a toroidal wiring consisting of four rectangular loops and analyzed its response in terms of multipole moments \cite{9}, however toroidal moment is not dominant in the structure. It was found toroidal metamaterials can make negative index of refraction and rotate polarization state of light \cite{8,9}. In 2010 toroidal dominated response in microwave region was first experimentally demonstrated by three-dimensionally arranging four split ring resonators (SRRs) in a unit cell of toroidal symmetry \cite{10}. Although there are some attempts in manufacturing three-dimensional SRR structures, it is of challenge not only at optical frequencies but also at microwave frequencies \cite{10}. Stacking of two-dimensional planar plasmonic metamaterial structures is assumed to be the pathway towards the third dimension in the optical regime. These metamaterials have exhibited exotic properties and potential applications: artificial magnetic composite \cite{11}, perfect absorber \cite{12}, electromagnetically induced transparency medium \cite{13,14,15,16}. Planar structure based scheme is not limited at microwave band, but also show performance at Tera-hertz band \cite{17} and even optical range \cite{18,19,20}.\begin{figure}[b]
\includegraphics[width=8.6cm]{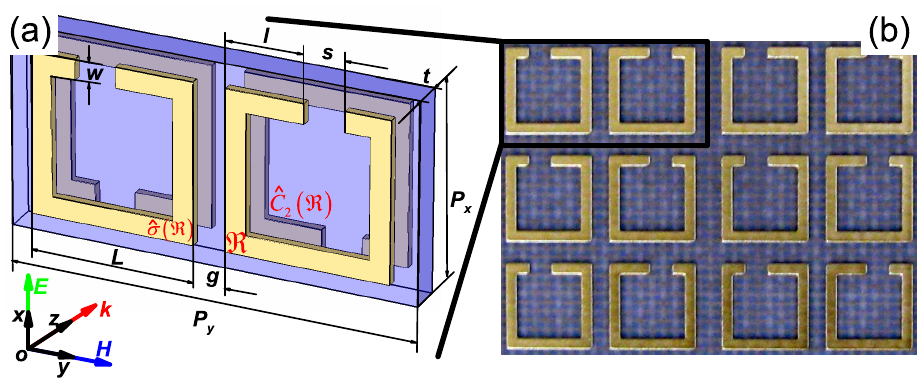}\caption{(a) Schematic of the unit cell for the ASRR-based planar toroidal metamaterial and corresponding electromagnetic excitation configuration (with polarization direction along $x$ axis). (b) Photograph of the toroidal metamaterial slab, geometric parameters are denoted by black letters in Fig. 1(a).}\label{fg1}
\end{figure}

In this paper, we present a planar scheme for toroidal metamaterial, with the unit cell comprised of four asymmetric split ring resonators (ASRRs), we show that a toroidal-molecule can be constructed through rational arrangement of planar ASRRs as meta-atoms via manipulating structural symmetry among the meta-atoms. Field maps clearly indicate that the toroidal resonance paves fantastical electromagnetic field confinement style in a subwavelength-sized head-to-tail channel of the toroidal geometry. Planar scheme by manipulating the coupling among the ASRRs may stimulate research in optical region involving toroidal multipoles. Toroidal geometry together with the Fano resonance made high-Q response will have enormous potential applications in low-threshold lasing, cavity quantum electrodynamics and nonlinear processing.

The designed planar structure is comprised of two metallic layers and a dielectric spacer layer. A unit cell of our model slab or the toroidal-molecule is shown stereographically in Fig.~\ref{fg1}(a), the toroidal-molecule is constructed through rational arrangement of four asymmetric split square ring resonators (ASRRs) as meta-atoms via manipulating structural symmetry of the meta-atoms. Pairs of ASRRs in the same layers are of mirror symmetry about the $xoz$ plane ($\sigma$); pairs of ASRRs in the different layers are of $180^\circ$ rotational symmetry about the $y$ axis ($C_2$). We note that throughout our study metamaterials are illuminated by $x$-polarized electromagnetic waves both in experiments and calculations as illustrated in Fig.~\ref{fg1}(a).

\begin{figure}[bp]
\includegraphics[width=8.6cm]{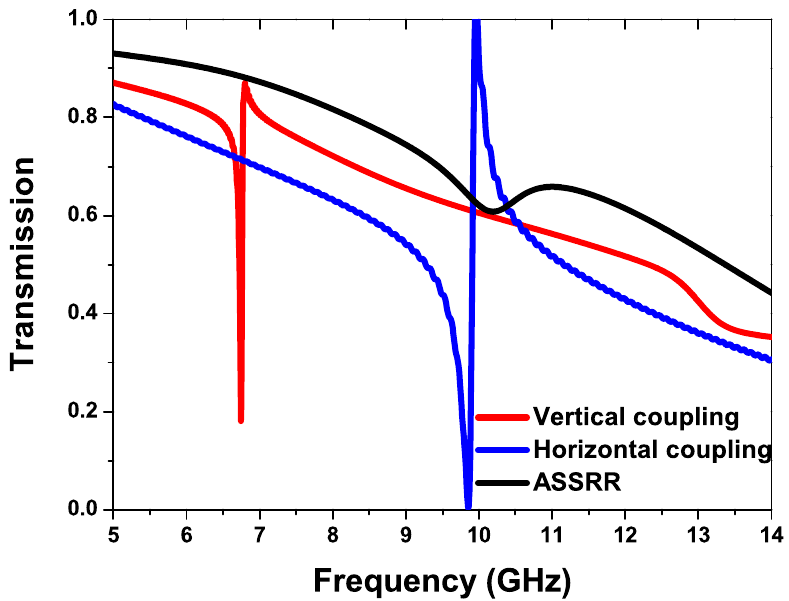}\caption{Comparison of FDTD calculated transmission spectra for planar metamaterial unit cell with only single ASRR (black curve), vertical coupled ASRR-pair (red curve) and horizontal coupled ASRR-pair (blue curve).}\label{fg2}
\end{figure}

Toroidal-molecule in our model system is constructed from four contiguous ASRRs as meta-atoms, since these meta-atoms are in close proximity, the nearest-neighbor mutual interactions or hybridizations \cite{21,22,23,24} between them play determinative role in their collaborative response to external excitations. We firstly investigate response of single ASRR and then mutual coupling interactions in pairs of the ASRRs. We performed all numerical calculations with a brute-force finite-difference-in-time-domain (FDTD) \cite{25} electromagnetic solver (Gallop) \cite{26}. For a metamaterial with its unit cell only one ASRR, it is showed that the \textit{trapped mode} can be excited even under normal electromagnetic incidence condition in a recent study \cite{27}, the trapped mode originates from symmetry breaking to some extent of the metallic element of a metamaterial. Small structural asymmetry introducing makes a degeneration from \textit{dark mode} \cite{28} to the trapped mode, or we can say weakly coupled nature to external excitation of ASRR is an inherited attribute from the dark mode. For the single ASRR made unit cell case, frequency plot performance of the transmission is shown as black solid line in Fig.~\ref{fg2}. The dip position near $10$ GHz indicate the trapped mode of ASRR, the resonant response appear to have Fano line-shape \cite{29,30,31}, result in destructive interference between trapped sharp resonance and free space propagating continuum-like spectrum. Meanwhile, we noted that the single ASRR is of weak excitation, again verify that the trapped mode of ASRR is weakly coupled to external excitation since electromagnetic wave can easily leak through the unpatterned area of the meta-surface.

Then we consider coupling interactions between the ASRRs. Numerical simulations were first carried out for the structure of in-plane ASRR pair [i.e., $\Re$ and $\hat{\sigma}(\Re)$ in Fig.~\ref{fg1}(a)] as fundamental element, the transmittance spectrum is represented by the blue solid curve in Fig.~\ref{fg2}, in which only one resonance is clearly observable. We call this kind of interaction as horizontal coupling, the single peak response comes from in-plane mirror symmetry, only mirror symmetric surface current can be excited under normal incidence because anti-symmetry surface current can not be excited since there is no phase retardation, the anti-symmetric mode is now a dark mode \cite{28}.

For the structure of stacked ASRR pair [i.e., $\Re$ and $\hat{C}_2(\Re)$ in Fig.~\ref{fg1}(a)] as fundamental element, the transmittance spectrum is represented by the red solid curve in Fig.~\ref{fg2}, there now exist two resonances. We call this kind of interaction as vertical coupling, the two resonances are far away from the original magnetic trapped mode of ASRR, which is due to extreme overlap of the two magnetic trapped modes of ASRRs in the vertical coupling configuration. We also note that for the low-frequency mode magnetic oscillations of the ASRRs are in-phase, this is crucial for toroidal response in our planar metamaterial system, the 2-fold rotational symmetry of the meta-molecule made the in-phase response at low frequency.

\begin{figure}[ptb]\includegraphics[width=8.6cm
]{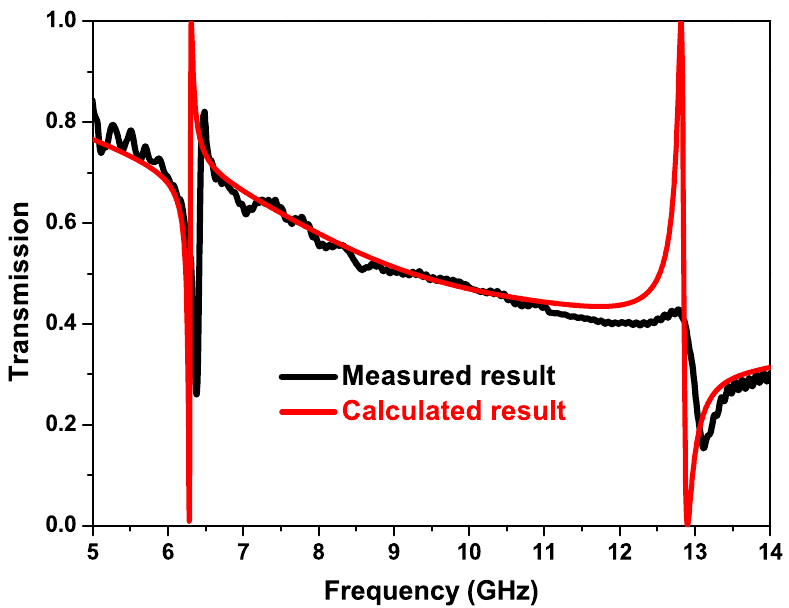}\caption{Experimentally measured (black curve) and FDTD calculated (red curve) dispersion of transmitted power of the designed planar toroidal metamaterial.}\label{fg3}
\end{figure}

We construct a meta-molecule involving previously mentioned horizontal coupling and vertical coupling with four ASRRs for toroidal response. Figure~\ref{fg3} presents the FDTD calculated and experimentally measured transmission spectra. The calculated intensity of transmitted wave (red solid line) shows clearly two Fano shaped resonances come from destructive interference between collective sharp response of the meta-molecule and free space propagating continuum-like spectrum.

\begin{figure}[t]
\includegraphics[width=7.6cm]{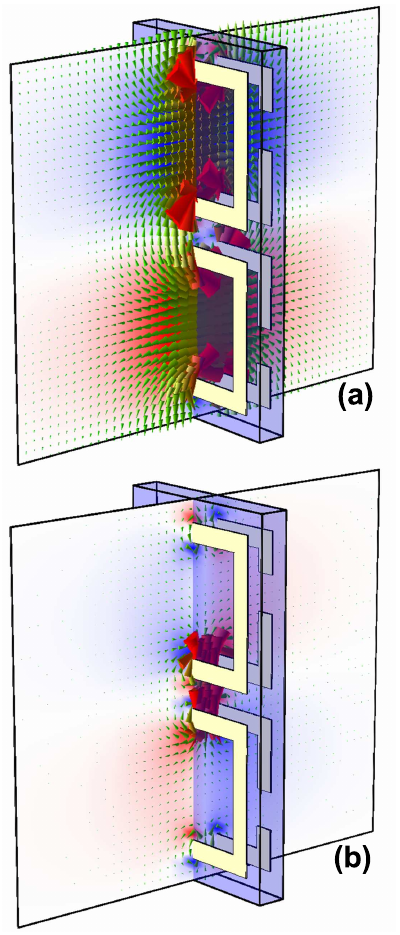}\caption{Calculated spatial distributions of vectorial magnetic field for correspondingly toroidal mode and magnetic mode plotted over the color maps of $z$-component magnetic field at peak frequencies 6.378GHz (a) and 12.905GHz (b) respectively.}\label{fg4}
\end{figure}

The two resonances have their characteristic frequency nearly the same as vertical coupled ASRR pair (red solid curve in Fig.~\ref{fg2}), frequency of the upper resonance at $12.905$ GHz (marked with the peak frequency) is determined by the out-phase vertical coupling, which is clearly observable as in the color map of $z$-component magnetic field in Fig.~\ref{fg4}(b), the coupling in this manner causes a net nonzero magnetic component along $y$ axis; meanwhile, frequency of the lower resonance at $6.378$ GHz (marked with the peak frequency) is determined by the in-phase vertical coupling, in-phase coupled magnetic mode [see color map of the $z$-component magnetic field in Fig.~\ref{fg4}(a)] together with horizontal coupling under normal electromagnetic incidence made a circumfluent magnetic field or magnetic votex, this is rightly a picture of the toroidal dipolar \cite{10} which shows a head-to-tail magnetic configuration.

We fabricated a sample with exactly the same parameters as our theoretical model. A printed circuit board (PCB) with metal sheets ($0.035$ mm-thick) attached to both sides of a $0.787$ mm-thick commercial Taconic TLX-$8$ dielectric substrate with the permittivity $\epsilon_r = 2.55$ is employed. The meta-molecule array has a lattice constant of $P_x = 5$ mm and $P_y = 10$ mm. The metal strip width is $w = 0.5$ mm wide, the outer width of the ASRRs are $L = 4$ mm, inner gap between the two in-plane ASRRs is $g = 0.8$ mm, and a $s = 1$ mm wide split is opened on the original square ring with a distance $l = 1.9$ mm away from the center sideline, overall lateral size of our planar metamaterial slab is $540$ mm $\times$ $440$ mm. The transmission spectra through our sample were measured with a vector network analyzer Agilent 8722ES in an anechoic chamber.

The measured results (black solid curve in Fig.~\ref{fg3}) meet quite well with the theoretical predictions. Two resonances of our model system are all clearly observable in the experimental spectrum. However we found a curious and abnormal phenomenon: the low-Q coupled magnetic resonance at high frequency is less obvious while the high-Q toroidal resonance at low frequency is quite obvious. We believe that this is due to the exotic field localization configuration of the toroidal geometry. This can be checked by looking at the evolution of magnetic field inside the structure as calculating spatial field distribution showed in Fig.~\ref{fg4}: in the toroidal mode (at $6.378$ GHz) case, most field is restricted to flowing inside the head-to-tail channel while for the coupled magnetic mode little field is confined inside the channel, corresponding to in-phase and out-phase vertical coupling as illustrated in the color maps of z-component magnetic field which show continuum or reversion. It can be see that strongly enhanced field is adhered and localized on metal surfaces in the coupled magnetic mode (at $12.905$ GHz) as in previous metamaterial structures, Planar toroidal metamaterial offers strikingly different style for electromagnetic is of convincing manifestation about the abnormal low loss phenomenon of the toroidal mode. The findings have beneficial for low Joule loss metallic photonic metamaterials \cite{32}.\begin{figure}[ptb]
\includegraphics[width=8.6cm]{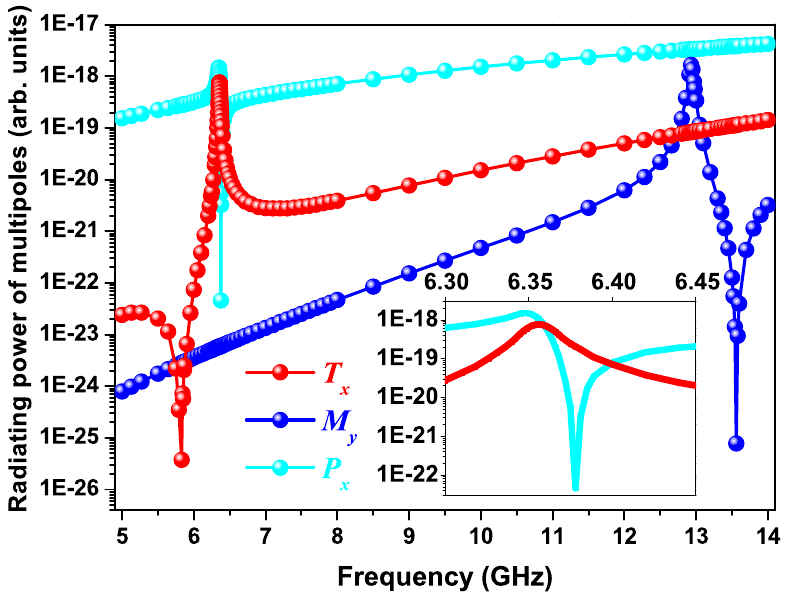}\caption{Frequency dependence of radiating power for multipoles from the planar toroidal metamaterial: toroidal dipole (red curve), magnetic dipole (blue curve) and electric dipole (cyan curve). The inset illustrates radiating strength of the toroidal dipole and electric dipole around 6.378GHz.}\label{fg5}
\end{figure}

Radiating power for induced multipole moments is calculated through integrating spatial distributed fields over volume of the planar meta-molecule with the current density formalization \cite{10}. Figure~\ref{fg5} shows the radiating power of three dominative multipoles of the planar metamaterial system: electric dipole (cyan curve), magnetic dipole (blue curve) and toroidal dipole (red curve). As can be seen from Fig.~\ref{fg5} electric dipole $P_x$ shows large components in the full frequency band, which reveal the fact that our planar metamaterial system is electric excited and the planar structure is hard to coupling with the incoming parallel magnetic component. Generally we notice that magnetic dipole $M_y$ and toroidal dipole $T_x$ show considerable scattering contributions at $6.356$ GHz and $12.930$ GHz which are just near the two transmission peak frequencies. Though $x$-component of electric dipole moment $P_x$ show strongest contribution at the $12.905$ GHz transmission peak frequency, the $y$-component of the magnetic dipole moment $M_y$ also provide significant contribution which can be checked in the time-domain simulated field patterns. While for the $6.378$ GHz transmission peak, the $x$-component of the toroidal dipole moment $T_x$ plays decisive role in comparison with electric dipole and magnetic dipole, inset in Fig.~\ref{fg5} clearly illustrates radiating strength of the toroidal dipole moment $T_x$ and electric dipole moment $P_x$ around $6.378$ GHz, we see that at $6.378$ GHz radiating power for the toroidal dipole moment is three orders more of magnitude larger than that for electric dipole moment, this is due to electric dipole radiation shows a Fano line-shaped spectrum that comes from radiating interference of induced current in the trapped mode. All these indicate that the experimentally observed transmission peak around $6.378$ GHz is the toroidal moment response of our planar metamaterial.

In summary a planar metamaterial structure comprised of ASRRs is proposed for observation of toroidal moment response. It is showed that a toroidal-molecule can be constructed through rational arrangement of planar ASRRs as meta-atoms via manipulating structural symmetry of the meta-atoms. The characteristic spectrum is experimentally verified at microwave range. Field maps and multipole moment calculations clearly indicate that the toroidal resonance paves new electromagnetic confinement style in a subwavelength scale. The planar toroidal structure can be generalized to terahertz, infrared, and visible frequencies particularly for low Joule loss metallic photonic metamaterials, the novel field localization configuration with a high-Q response and low loss have beneficial for myriad photonic applications, e.g., low-threshold lasing, cavity quantum electrodynamics and nonlinear processing.

\begin{acknowledgments}
This work was supported by NSFC (Grant Nos. 11174221 and 10974144), CNKBRSF (Grant No. 2011CB922001), the National 863 Program of China (Grant No. 2006AA03Z407), NCET (Grant No. 07-0621), STCSM and SHEDF (Grant No. 06SG24). Work at Ames Laboratory was supported by the Department of Energy (Basic Energy Sciences, Division of Materials Sciences and Engineering) under Contract No. DE-AC02-07CH11358. The author Y. Fan acknowledges financial support from China Scholarship Council (No. 201206260055).
\end{acknowledgments}

\end{document}